\documentclass[prd,aps,showpacs,nofootinbib,preprint]{revtex4-1}
\usepackage{graphicx,color,amsmath,amsxtra}
\usepackage{epsf}
\usepackage{amssymb}
\usepackage{enumerate}
\usepackage{hhline}
\usepackage{array}
\usepackage{tabularx}
\usepackage{hangcaption}

\newcommand{\newc}{\newcommand}
\newc{\be}{\begin{equation}}
\newc{\ee}{\end{equation}}
\newc{\beq}{\begin{eqnarray}}
\newc{\eeq}{\end{eqnarray}}

\begin{document}

\rightline{\tt{MIT-CTP-4231}}

\title{Alternative conformal quantum mechanics}

\author{S.-H. Ho\footnote{E-mail address: shho@mit.edu}}

\affiliation{Center for Theoretical Physics \\
Massachusetts Institute of Technology \\
 Cambridge, MA 02139}

\begin{abstract}

We investigate a one dimensional quantum mechanical model, which is invariant under translations and dilations but does not respect the conventional conformal invariance. We describe the possibility of modifying the conventional conformal transformation such that a scale invariant theory is also invariant under this new conformal transformation.

\end{abstract}

\maketitle

It is known that a scale invariant theory does not always enjoy conventional conformal symmetry, where the configuration variables transform in the conventional way, rendering them to be primary operators (invariant under conformal transformations at the origin)~\cite{Jackiw:2011vz, ElShowk:2011gz}. We examine these issues for a theory in one time dimension, i.e. quantum mechanics for a single quantum variable $q(t)$.

The theory is governed by a one-dimensional Lagrangian $L=L(q, \dot{q}; t)$. 
When it takes the form
\beq \label{eq1}
L=f(x) q^{-2},
\eeq
where $x \equiv \dot{q}q$ and $f(x)=x^n$ with a real number $n$ \footnote{The form of Lagraingian}, it enjoys time translation and scale invariance
\begin{subequations} \label{eq2}
\beq
\label{eq2a}& &\ \  \delta_T q =\dot{q} \\
\label{eq2b}& &\ \  \delta_D q = t \dot{q} - \frac{1}{2}q 
\eeq
but is conventionally conformally invariant
\beq
\label{eq2c}& &\ \  \delta_C q =t^2 \dot{q}- t q 
\eeq
\end{subequations}
only when~\cite{Jackiw:1972cb,de Alfaro:1976je, Jackiw:1980mm}
\beq \label{eq3}
L=\frac{1}{2}\dot{q}^2 - \frac{\lambda} {2q^{2}} . 
\eeq
The associated conserved quantities can be obtained by Noether's theorem 
\begin{subequations} \label{eq4}
\beq 
\label{eq4a} H && = \frac{1}{2}p^2 + \frac{\lambda} {2q^{2}} , \ \ p\equiv \frac{\partial L}{\partial \dot{q}}=\dot{q} \\
\label{eq4b} D && =tH - \frac{1}{4}(q\dot{q}+\dot{q}q)=tH-\frac{1}{4}(qp+pq) \\
\label{eq4c} K && =-t^2 H +2tD +\frac{q^2}{2}.
\eeq
\end{subequations}

In this paper we describe the possibility of changing the conformal transformation law (\ref{eq2c}) for $q$, such that conformal invariance holds for the entire family of Lagrangians (\ref{eq1}). Time translation and dilatations remain conventional, leading to constants of motion
\begin{subequations} \label{eq5}
\beq 
\label{eq5a} H = && p\dot{q}-f(x)q^{-2}  \\
\label{eq5b} D = && tH - \frac{1}{4}(qp+pq) .
\eeq
\end{subequations}
But the conformal generator is changed\footnote{One may consider the more general case where $q$ has an arbitrary scale dimension $d$. Then a scale invariant Lagrangian becomes $L'=\dot{q}^n q^{\frac{1-n}{d}-n}$. This provides a complicated generalization which I have not studied.}.

To begin, we assume that a conserved conformal generator $K$ exists and together with $H$ and $D$ generates the group $SO(2,1)$:
\begin{subequations} \label{eq6}
\beq 
\label{eq6a} i \left[D, H \right]= && + \hbar H, \\
\label{eq6b} i \left[D, K \right]= && - \hbar K, \\
\label{eq6c} i \left[K, H \right]= &&  2  \hbar D. 
\eeq
\end{subequations}
The Casimir operator $\mathcal{C}$ of $SO(2,1)$ is 
\beq \label{eq7}
\mathcal{C}=\frac{1}{2}(KH+HK)-D^2 .
\eeq
From (\ref{eq6}) and (\ref{eq7}), we can derive the general form of conformal generator $K$ for given $H$ and $D$ :
\beq \label{eq8}
K=&& -H^{-1} K H +2 H^{-1} D^2 +2H^{-1} \mathcal{C} \nonumber \\
=&& -\left(\left[H^{-1}, K\right]+KH^{-1}\right)H+2 \left(\left[H^{-1}, D\right]+DH^{-1}\right)D+2H^{-1} \mathcal{C} \nonumber \\
=&&-\left( -H^{-1}(2 i \hbar D)H^{-1}+KH^{-1}\right)H+2 \left(-H^{-1}(i \hbar )+D H^{-1}\right)D+2H^{-1} \mathcal{C} \nonumber \\
=&& 2i\hbar H^{-1}D-K-2i\hbar H^{-1}D+2D H^{-1}D+2 H^{-1}\mathcal{C} \nonumber \\
=&& -K +2 D H^{-1} D+2 H^{-1} \mathcal{C}
\eeq
Therefore, the expression of $K$ is ~\cite{ref5}
\beq \label{eq9}
K=D H^{-1}D + \mathcal{C}H^{-1}.
\eeq
%
The addition to $K$ of the $\mathcal{C}/H$ term reflects the fact that the $SO(2,1)$ (\ref{eq6}) algebra is unchanged when the generators are modified according to
\beq 
H && \rightarrow H+\frac{a}{K} \nonumber \\
K && \rightarrow K+\frac{b}{H} \nonumber 
\eeq
provided $ab=0$. We allow for the conformal modification, so we set $a=0$. Because of this ambiguity, $\mathcal{C}$ parametrizes an entire family of $K$.

%
%

First we consider the conventional case $L_2=\frac{\dot{q}^2}{2}$ and $H=\frac{p^2}{2}$ (the interacting case $H=\frac{p^2}{2}+\frac{\lambda}{2q^2}$ is too difficult to treat quantum mechanically owing to $H^{-1}$ in (\ref{eq9})). We can write the dilatation generator (\ref{eq5b}) and conformal generator (\ref{eq9}) as 
\beq 
\label{eq10} D=&& tH - \frac{1}{4}(qp+pq) = tH+D_0 \\
\label{eq11} K=&&DH^{-1}D+\frac{\mathcal{C}}{H}  =(tH+D_0)H^{-1}(tH+D_0)+\frac{\mathcal{C}}{H} \nonumber \\
=&&t^2 H+2 t D_0+D_0 H^{-1} D_0 +\frac{\mathcal{C}}{H} = - t^2 H+2tD +K_0 
\eeq
where $D_0$ and $K_0$ are the generators at $t=0$:
\beq
\label{eq12}  D_0 && \equiv - \frac{1}{4}(qp+pq) \\
\label{eq13}  K_0 && \equiv D_0 H^{-1} D_0 + \frac{\mathcal{C}}{H}.
\eeq 
From (\ref{eq13}), we can derive $K_0$:
\beq \label{eq14}
K_0=\frac{q^2}{2}+\frac{3\hbar^2}{8}\frac{1}{p^2}+ \frac{\mathcal{C}}{H}= \frac{q^2}{2}+\left(\frac{3\hbar^2}{16}+\mathcal{C}\right) \frac{1}{H}.
\eeq
In order for $q$ to be a primary operator, we choose $\mathcal{C}=-\frac{3 \hbar^2}{16}$ and $K_0=\frac{q^2}{2}$. 
The transformation law of $q$ in (\ref{eq2}) can be reproduced by commutation and the usual formulas emerge:
\begin{subequations}\label{eq15}
\beq 
\label{eq15a} &&  \delta_T q =\frac{i}{\hbar} \left[ H, q \right] = p=\dot{q} \\
\label{eq15b} &&  \delta_D q =\frac{i}{\hbar} \left[ D, q \right] = t \dot{q} - \frac{1}{2}q \\
\label{eq15c} &&  \delta_C q =\frac{i}{\hbar} \left[ K, q \right] = t^2 \dot{q} - t q    
\eeq
\end{subequations}

%
%

Next we derive $K$ for the special case $n=4/3$. 
We show in Appendix \ref{Appendix A}, where we examine arbitrary $n$, that $n=4/3$ is the only other case with a primary operator. We start from the Lagrangian
\beq \label{eq16}
L_{4/3} =\frac{3}{4}\dot{q}^{4/3}q^{-2/3}
\eeq
From (\ref{eq5}), we have the classical Hamiltonian 
\beq \label{eq16-1}
H=\frac{1}{4}p^4 q^2, \ \ p=\dot{q}^{1/3}q^{-2/3}.
\eeq
We then write (\ref{eq16-1}) and dilatation generators in a quantum mechanical way with the ordering\footnote{There are various ways to order the Hamiltonian as a Hermitian operator. Certainly different orderings give different $K$ and Casimir $\mathcal{C}$, but will result in the same classical transformation  law $\delta_C q$. In the Appendix \ref{Appendix B}, we will consider another ordering of $H= \frac{1}{4}  q p^4 q$.}:
\begin{subequations} \label{eq17}
\beq
\label{eq17a} H=&& \frac{1}{4}  p^2 q^2 p^2 , \\
\label{eq17b} D=&& tH-\frac{1}{4}(pq+qp)\equiv tH+D_0 
\eeq
\end{subequations}
From (\ref{eq13}), $K_0$ is given by
\beq \label{eq18}
K_0 && = D_0 H^{-1} D_0+\frac{\mathcal{C}}{H}  \nonumber \\
=&& p^{-2}+ \left( \frac{3 \hbar^2}{16}+\mathcal{C}\right) \frac{1}{H}
\eeq
%
By choosing Casimir $\mathcal{C}= - \frac{3 \hbar^2}{16}$ we can eliminate the $H^{-1}$ term in $K_0$ and $K$ (note the standard case also has $\mathcal{C}= - \frac{3 \hbar^2}{16}$):
\begin{subequations} \label{eq19}
\beq
\label{eq19a} K_0 = &&  p^{-2} \\
\label{eq19b} K = && -t^2 H +2 t D +  p^{-2} . 
\eeq
\end{subequations}
%
Since $\left[ K_0, p \right]=0$, the primary operator in the present case is canonical momentum $p$. 

We now have the explicit expressions of $H$, $D$ and $K$ so we can reproduce the transformation law for $q$:
\begin{subequations} \label{eq20}
\beq \label{eq20a}
\delta_T q =&&\frac{i}{\hbar} \left[ H, q \right]=\frac{1}{2}\ \left( p^2 q^{2} p + p  q^{2} p^2 \right) \equiv \dot{q}  \\
\label{eq20b}
\delta_D q = && \frac{i}{\hbar} \left[ tH+D_0, q \right]=
t \dot{q} -\frac{1}{2}q\\
\label{eq20c}
\delta_C q =&& \frac{i}{\hbar} \left[ - t^2H+2tD+K_0, q \right] \nonumber \\
=&&
t^2 \dot{q}-t q-2  p^{-3} 
\eeq
Substituting (\ref{eq16-1}) into (\ref{eq20c}), we obtain the classical transformation law for $q$:
\beq \label{eq20d}
\delta_C q = t^2 \dot{q} -tq -2 \dot{q}^{-1} q^2
\eeq
\end{subequations}

%
%
%

Next we derive the classical form of the new conformal constant of motion by applying Noether's theorem to the Lagrangian (\ref{eq16}). The variation of the Langrangian under (\ref{eq20d}) can be written as a total time-derivative without using equations of motion:
\beq \label{eq22}
\delta_C L_{4/3} && =\frac{\partial L_{4/3}}{\partial \dot{q}} \delta_C \dot{q} + \frac{\partial L_{4/3}}{\partial q} \delta_C q \nonumber \\
&& =\frac{d}{dt} (t^2\dot{q}^{4/3}q^{-2/3}-4 \dot{q}^{-2/3}q^{4/3}) \equiv \frac{d}{dt} X
\eeq
which indicates that (\ref{eq20c}) is a symmetry transformation of $L_{4/3}$ (\ref{eq16}). The constant of motion is 
\beq \label{eq23}
K=\frac{\partial L_{4/3}}{\partial \dot{q}} \delta_C q - X 
= && \frac{t^2}{4}  p^4 q^2 - tq + p^{-2}.
\eeq
This is the classical version of (\ref{eq19b}).


\appendix

\section{Conformal generator for arbitrary $n>1$} \label{Appendix A}

In this appendix, we derive the conformal generator $K$ for Lagrangian in (\ref{eq1})
\beq \label{eqL}
L=\frac{1}{n}\dot{q}^nq^{n-2}
\eeq
for arbitrary $n>1$ and two different orderings of the Hamiltonian $H$ which have the same classical transformation law of $q$ with different Casimirs $\mathcal{C}$. 

\subsection{$H=\frac{N}{N+1} p^{\frac{1}{2}\left(1+\frac{1}{N}\right)}q^{-1+\frac{1}{N}} p^{\frac{1}{2}\left(1+\frac{1}{N}\right)}$, $N \equiv n-1$}
We start from the classical Hamiltonian $H$:
\begin{subequations}
\beq \label{eqB0}
\label{eqB0a}  H=&&\frac{N}{N+1} p^{1+\frac{1}{N}} q^{-1+\frac{1}{N}}   , \\
\label{eqB0b}  p=&&  \dot{q}^N q^{N-1}.
\eeq
\end{subequations}
Then we write (\ref{eqB0}) and time dilatation operator $D$ associated with Lagrangian (\ref{eqL}) quantum mechanically with the ordering
%
\begin{subequations} \label{eqB1}
\beq 
\label{eqB1a} H=&&
		\frac{N}{N+1} p^{\frac{1}{2}\left(1+\frac{1}{N}\right)} q^{-1+\frac{1}{N}} p^{\frac{1}{2}\left(1+\frac{1}{N}\right)} \\ 
\label{eqB1b} D=&& t H - \frac{1}{4}(pq+qp) \equiv t H+D_0, 
\eeq
and using (\ref{eq11}) for conformal operator $K$: 
\beq
\label{eqB1c}
K
=&&t^2 H+2tD_0 +K_0. 
\eeq
%
\end{subequations}
After a straightforward calculation, the $K_0$ is:
\beq \label{eqB2}
K_0&& = D_0 H^{-1} D_0 + \frac{\mathcal{C}}{H}   \nonumber \\
=&&\frac{(N+1)}{4N} \ p^{\frac{1}{2}\left(1-\frac{1}{N}\right)} q^{3-\frac{1}{N}} p^{\frac{1}{2}\left(1-\frac{1}{N}\right)}+\left( \frac{ \hbar^2 } {16} \left(\frac{4}{N}-\frac{1}{N^2}\right)+\mathcal{C}\right) \frac{1}{H}.
\eeq
By choosing Casimir $\mathcal{C}$
\beq
\label{eqB4} \mathcal{C}=&& - \frac{ \hbar^2 } {16} \left(\frac{4}{N}-\frac{1}{N^2}\right),  
\eeq
we eliminate the $H^{-1}$ term in $K_0$:
\beq 
\label{eqB3} K_0=
&&\frac{(N+1)}{4N} \ p^{\frac{1}{2}\left(1-\frac{1}{N}\right)} q^{3-\frac{1}{N}} p^{\frac{1}{2}\left(1-\frac{1}{N}\right)}.
\eeq
Note from (\ref{eqB3}) that only $n=2$ $(N=1)$ and $n=4/3$ $(N=1/3)$ produce a primary operator.

The transformation law for $q$ can be reproduced by computing
\begin{subequations} \label{eqB5}
\beq \label{eqB5a}
\delta_T q =&&\frac{i}{\hbar} \left[ H, q \right]
\nonumber \\
=&&\frac{1}{2} \left( p^{\frac{1}{2}\left(1+\frac{1}{N}\right)} q^{-1+\frac{1}{N}} p^{\frac{1}{2}\left(-1+\frac{1}{N}\right)}+p^{\frac{1}{2}\left(-1+\frac{1}{N}\right)} q^{-1+\frac{1}{N}} p^{\frac{1}{2}\left(1+\frac{1}{N}\right)}\right)  \equiv \dot{q} \\
\label{eqB5b}
\delta_D q = && \frac{i}{\hbar} \left[ tH+D_0, q \right] 
= t\dot{q}-\frac{1}{2}q \\
\label{eqB5c}
\delta_C q =&& \frac{i}{\hbar} \left[ t^2H+2tD_0+K_0, q \right] \nonumber \\
=&& t^2 \dot{q}
-tq +\frac{N^2-1}{8N^2} p^{-\frac{1}{2}\left(1+\frac{1}{N}\right)}\left(p q^{3-\frac{1}{N}}+q^{3-\frac{1}{N}} p\right) p^{-\frac{1}{2}\left(1+\frac{1}{N}\right)}  
\eeq
We then obtain the classical transformation law of $q$ by using (\ref{eqB0b}):
\beq
\label{eqB5d} \delta_C q = && t^2 \dot{q}-tq+ \frac{1}{4} \left(1-\frac{1}{N^2}\right) \dot{q}^{-1}q^2 
\eeq
\end{subequations}
%
%
%
%

\subsection{$H=\frac{N}{N+1} q^{\frac{1}{2}\left(-1+\frac{1}{N}\right)} p^{1+\frac{1}{N}}q^{\frac{1}{2}\left(-1+\frac{1}{N}\right)}$ }
We change (\ref{eqB1a}) to
\beq \label{eqB7}
H=
\frac{N}{(N+1)} q^{\frac{1}{2}\left(-1+\frac{1}{N}\right)} p^{1+\frac{1}{N}} q^{\frac{1}{2}\left(-1+\frac{1}{N}\right)}  .
\eeq
%
%
The generator $K_0$ is:
\beq \label{eqB8}
K_0 && =D_0 \frac{1}{H} D_0+\frac{\mathcal{C}}{H} \nonumber \\
=&&\frac{N+1}{4N} \ q^{\frac{1}{2}\left(3-\frac{1}{N}\right)}  p^{1-\frac{1}{N}}  q^{\frac{1}{2}\left(3-\frac{1}{N}\right)} +\frac{\hbar^2}{16}\left( \left(4-\frac{1}{N^2}\right)+\mathcal{C}\right) \frac{1}{H} .
\eeq
The choice of Casimir $\mathcal{C}$ 
\beq \label{eqB9}
\mathcal{C}= - \frac{\hbar^2}{16}\left(4-\frac{1}{N^2}\right) ,
\eeq
leads to
\beq \label{eqB10}
 K_0=\frac{N+1}{4N} \ q^{\frac{1}{2}\left(3-\frac{1}{N}\right)}  p^{1-\frac{1}{N}}  q^{\frac{1}{2}\left(3-\frac{1}{N}\right)}.
\eeq

%
%
%
Following the same procedure, we obtain the classical transformation law of $q$
\begin{subequations} \label{eqB11}
\beq
\label{eqB11a} \delta_T q =&& \dot{q} \\
\label{eqB11b} \delta_D q = && t\dot{q}-\frac{1}{2}q \\
\label{eqB11c} \delta_C q = && t^2 \dot{q}-tq+ \frac{1}{4} \left(1-\frac{1}{N^2}\right) \dot{q}^{-1}q^2 
\eeq
\end{subequations}
which are the same as (\ref{eqB5}). 

We can see the variation of Lagrangian (\ref{eqL}) under transformation (\ref{eqB11c}) 
\beq \label{eqB12}
\delta_C L=&&\frac{\partial L}{\partial \dot{q}} \delta_C \dot{q} + \frac{\partial L}{\partial q} \delta_C q \nonumber \\
=&&\frac{d}{dt}\left(t^2 L - \frac{(N+1)^2}{4N^2} \dot{q}^{N-1}q^{N+1} \right)
\eeq
is a total time-derivative. The constant of motion is
\beq \label{eqB13}
K=&&\frac{\partial L}{\partial \dot{q}} \delta_C q -\left(t^2 L - \frac{(N+1)^2}{4N^2} \dot{q}^{N-1}q^{N+1} \right) \nonumber \\
=&&  \frac{t^2 N}{N+1} p^{1+\frac{1}{N}} q^{-1+\frac{1}{N}} -tpq + \frac{N+1}{4N} p^{1-\frac{1}{N}}q^{3-\frac{1}{N}}
\eeq
This is classical form of (\ref{eqB1c}) with $K_0$ given in (\ref{eqB3}) (and (\ref{eqB10})).


\section{Another ordering of Hamiltonian for $n=4/3$} \label{Appendix B}

In this appendix, we use for $H$ an expression, which is ordered differently from (\ref{eq17a}):
\beq \label{eqA1}
H=\frac{1}{4}  q p^4 q.
\eeq
Using (\ref{eq13}), we obtain
\beq \label{eqA2}
K_0=  p^{-2} + \left(\frac{-5 \hbar^2}{16}+\mathcal{C}\right)\frac{1}{H} .
\eeq
Hence, we choose Casimir to be
\beq \label{eqA4}
\mathcal{C}=\frac{5 \hbar^2}{16}.
\eeq
and have
\beq \label{eqA3}
K_0=  p^{-2} .
\eeq

Comparing (\ref{eq19a}) with (\ref{eqA3}), we observe the two $K_0$'s are the same in the two orderings. The Casimirs differ because of reordering.

\begin{acknowledgements}
I am grateful to Professor Roman Jackiw for providing to me the main idea of  this work
and continuous and patient guidance. This work is supported by the National Science Council of R.O.C. under Grant number:
NSC98-2917-I-564-122.

\end{acknowledgements}


\end{document}